%% file: main.tex
\documentclass[conference]{IEEEtran}
\IEEEoverridecommandlockouts
\usepackage{cite}
\usepackage{amsmath,amssymb,amsfonts}
\allowdisplaybreaks
\usepackage{algorithmic}
\usepackage{graphicx}
\graphicspath{{./}{./Figures/}}
\usepackage{textcomp}
\usepackage{xcolor}
\usepackage{fancyhdr}
\usepackage{subfig}

\usepackage{algorithm, algorithmic}

\usepackage{tikz}
\usetikzlibrary{arrows.meta,calc,positioning}
\usepackage{mathdots}
\usepackage{yhmath}
\usepackage{cancel}
\usepackage{color}
\usepackage[binary-units=true]{siunitx}
\DeclareSIUnit \VAr {VAr} 
\DeclareSIUnit \VA {VA} 
\DeclareSIUnit \rad {Radians} 

\usepackage{etoolbox}

\usepackage[justification=justified,font=footnotesize,skip=0pt]{caption}
\makeatletter
\let\old@ps@headings\ps@headings
\let\old@ps@IEEEtitlepagestyle\ps@IEEEtitlepagestyle
\def\psccfooter#1{%
    \def\ps@headings{%
        \old@ps@headings%
        \def\@oddfoot{\strut\hfill#1\hfill\strut}%
        \def\@evenfoot{\strut\hfill#1\hfill\strut}%
    }%
    \def\ps@IEEEtitlepagestyle{%
        \old@ps@IEEEtitlepagestyle%
        \def\@oddfoot{\strut\hfill#1\hfill\strut}%
        \def\@evenfoot{\strut\hfill#1\hfill\strut}%
    }%
    \ps@headings%
}
\makeatother

\def\BibTeX{{\rm B\kern-.05em{\sc i\kern-.025em b}\kern-.08em
    T\kern-.1667em\lower.7ex\hbox{E}\kern-.125emX}}
\begin{document}

\title{Fair Dynamic Operating Envelopes using Distributed Multi-Period Optimal Power Flow and Jain Index for Active Distribution Networks\\
{%
}
\thanks{This work was financially supported by the Brazilian National Council for Scientific and Technological Development — CNPq, under project grant 447271/2024-5.}
}

\author{\IEEEauthorblockN{Pedro Salomão Quessongo}
\IEEEauthorblockA{\textit{Dept. of Electrical Engineering} \\
\textit{Federal University of Paraná (UFPR)}\\
Curitiba, Brazil \\
pedro.quessongo@ufpr.br}
\and
\IEEEauthorblockN{Daniel Gebbran}
\IEEEauthorblockA{\textit{Dept. of Electrical Engineering} \\
\textit{Federal University of Paraná (UFPR)}\\
Curitiba, Brazil \\
daniel.gebbran@ufpr.br}
\and
\IEEEauthorblockN{Clodomiro Unsihuay-Vila}
\IEEEauthorblockA{\textit{Dept. of Electrical Engineering} \\
\textit{Federal University of Paraná (UFPR)}\\
Curitiba, Brazil \\
clodomiro.vila@ufpr.br}
\and

}

\makeatletter
\patchcmd{\@maketitle}
  {\addvspace{0.5\baselineskip}\egroup}
  {\addvspace{-1.0\baselineskip}\egroup}
  {}
  {}
\makeatother

\setlength{\textfloatsep}{2pt plus 1.0pt minus 1.0pt}
\setlength{\floatsep}{2pt plus 1.0pt minus 1.0pt}


\maketitle

\begin{abstract}
Dynamic operating envelopes (DOEs) are increasingly used to publish time-varying export limits that keep distribution networks within operational limits. Purely technical DOE allocation, however, can systematically privilege electrically favorable prosumers, while embedding fairness directly into a single-period optimal power flow (OPF) objective mixes network feasibility, equity and efficiency in a way that obscures the cost of fairness. This paper proposes a two-stage, multi-period framework that addresses both of these. Initially, a technical distributed OPF computes network-feasible export envelopes. The subsequent stage then applies a dynamic aggregate export budget and redistributes capacity through cumulative proportional fairness, limiting the additional curtailment by an admissible efficiency budget. The resulting fair DOEs are treated as first-stage decisions, while battery storage provides scenario-dependent recourse under demand and renewable uncertainty. The operational problem is solved by a calibrated regional alternating direction method of multipliers (ADMM) on a lossless LinDistFlow model and independently validated using AC power flow. On the IEEE~33-bus feeder over a 24-hour horizon, the technical benchmark yields \textcolor{black}{$\SI{2.0898}{\mega\watt\hour}$} of renewable curtailment, whereas the fairness-constrained allocation increases curtailment to \textcolor{black}{$\SI{5.7741}{\mega\watt\hour}$} but caps the maximum cumulative curtailment ratio at \textcolor{black}{$\SI{11.31}{\%}$} and raises Jain fairness indices close to unity, with AC voltage deviations below 0.01~p.u.\ and no voltage or thermal violations under the adopted 0.90--1.05~p.u.\ limits. Results show that considering both storage (which alleviates curtailment impact) and multi-period fairness (which increases curtailment) is an interesting approach for modern DOE design, which in turn requires a multi-period, co-designed approach.

\end{abstract}



\section{Introduction}
The increasing penetration of distributed energy resources (DERs), including photovoltaic (PV) generation, wind generation and battery energy storage systems (BESS), is changing distribution networks from passive feeders into active systems with bidirectional power flows \cite{REN21_2025}, which requires operational tools that can simultaneously preserve feeder integrity, reduce renewable curtailment and allocate export capacity fairly among prosumers \cite{Johanna_Mathieu_2025_IEEE_TEM__DR_DER}. The optimal power flow (OPF) is one such tool, which allows for conditioning the location and magnitude of distributed injections and assess their effects on voltage magnitudes, branch loading, thermal limits, substation power exchange and others \cite{molzahn2017}. However, DERs on distribution networks are too granular, and for a pure OPF strategy to work it would require micro-management, a capability which may not be desired by participants and often, meets technological barriers \cite{Chatzivasileiadis_2023_MicroFlex}. 

Dynamic Operating Envelopes (DOEs) address this challenge by defining time-varying export limits for prosumers while preserving distribution-network integrity \cite{lankeshwara2025}. When envelopes are computed from technical constraints alone, however, electrically favorable locations tend to receive larger export shares, so renewable curtailment becomes unevenly distributed among prosumers. Fairness is therefore an important complementary requirement in DOE allocation \cite{wickramasinghe2025}. At the same time, BESS and other flexible resources couple successive operating intervals through the state of charge, which means that a fair allocation defined independently at each time step may still accumulate uneven curtailment over the operating horizon. These observations motivate a multi-period DOE framework in which network-feasible technical envelopes, an explicit fairness stage, and storage recourse, are treated as distinct but coordinated decisions.

\subsection{Literature Review}

Because DOEs are commonly obtained from network-constrained optimization, the underlying power flow representation is a first design choice \cite{lankeshwara2025}. Radial distribution OPF is commonly represented through DistFlow or branch-flow models, which capture downstream power balances and voltage drops in a computationally tractable way \cite{baran1989}, offering a speed up against the traditional AC OPF for time-sensitive operations. Convex relaxations and linearizations of the branch-flow model further support operational studies on radial feeders \cite{farivar2013}. Nonetheless, even if linear power flow models remain useful approximations for multiphase distribution networks, their dispatch should be checked against nonlinear AC feasibility, at least in research and development stages \cite{bernstein2017}. 

Building on these OPF models, DOEs have emerged as a network-aware alternative to static export limits and to direct micro-management of behind-the-meter assets, publishing time-varying import and export capacities at customer connection points \cite{petrou2021}. OPF-based operating envelopes have been used to facilitate residential DER services while preserving network integrity without direct utility control of customer devices \cite{liu2022oe}. Recent reviews summarize DOE computation methods, allocation mechanisms and implementation challenges in distribution networks \cite{wickramasinghe2025}, and a complementary review places DOEs within the broader landscape of network-aware DER control, including centralized and distributed architectures \cite{lankeshwara2025}.

Once technical envelopes can be computed from OPF, a second issue arises. The export capacity shared among prosumers can be unfair. As such, fairness is a recurring concern because radial feeders amplify voltage sensitivity at electrically remote buses. A known quantitative fairness index, the Jain's fairness index, has been widely used to evaluate equality in resource allocation \cite{jain1984}. Furthermore, fair operating envelopes under uncertainty have been obtained through chance-constrained OPF \cite{yi2022}, and weighted proportional fairness has also been embedded in day-ahead operating-envelope OPF for low-voltage (LV) prosumers \cite{petrou2020fair}. Fairness-aware PV generation limits for voltage regulation have also been studied using conservative linear approximations \cite{gupta2024}, and technical and equitable DOE allocation perspectives have been compared in a two-stage MV--LV setting, confirming the trade-off between aggregate export capacity and disparity among connection points \cite{alam2023}.

These fairness-aware formulations are typically posed for a single interval, or embed equity directly in the OPF objective. Multi-period formulations become necessary when storage couples successive intervals through the state of charge. For example, \cite{gebbran2021} formulates a multi-period distributed OPF (DOPF) with fair DER coordination with Volt-Var control and PV curtailment, further showing that different equity principles produce distinct curtailment patterns. Moreover, distributed multi-period three-phase OPF with temporal neighbors have been shown to illustrate how temporal coupling can be exploited in decomposed network optimization \cite{pinto2020}. It has also been shown that flexible demand can likewise act as a recourse mechanism under stressed operating conditions \cite{li2011}. 

Because multi-period DER problems grow quickly with feeder size and the number of controllable assets, often generating mixed-integer (linear or non-linear) problems, distributed solution methods are attractive \cite{gebbran2023}. The alternating direction method of multipliers (ADMM) provides a standard framework for decomposed optimization with consensus constraints \cite{boyd2011}, and has been applied to distributed OPF on radial networks \cite{peng2018}. Broader distributed optimization and control methods for power systems are surveyed in \cite{molzahn2017}. 

Across this literature, the modeling, DOE, fairness and distributed multi-period strands remain only loosely coupled: fairness is typically embedded in a single allocation objective and assessed mainly on a per-interval basis, while technical and equitable DOE perspectives are often compared without an explicit multi-period budget that separates network-feasible envelopes from cumulative fairness and storage recourse.

\subsection{Contributions}
Against this background, the research gap addressed in this paper is the lack of a multi-period DOE workflow that separates network-feasible technical envelopes from an explicit budgeted fairness stage, so that the equity--efficiency trade-off can be quantified under storage coupling and distributed OPF coordination. The main contributions of this work are:
\begin{itemize}
    \item A two-stage DOE computation that first obtains technical export envelopes from network voltage and thermal limits, then redistributes capacity through cumulative proportional fairness under a dynamic aggregate export budget with an admissible curtailment-efficiency limit.
    \item A multi-period operational scheme in which the fair DOEs are first-stage decisions, with BESS providing scenario-dependent recourse under demand and renewable uncertainty.
    \item A calibrated regional ADMM solution of the resulting LinDistFlow DOPF on a partitioned radial feeder, together with independent nonlinear AC validation of operational feasibility for every operating period.
\end{itemize}
The framework is demonstrated on the IEEE~33-bus feeder, partitioned into nine regions over a 24-hour horizon, with quantitative comparison between the technical DOE benchmark and the fairness-constrained allocation, over different operating scenarios.

\section{Mathematical Formulation}

\begin{figure}[!tp]
    \centering
    \resizebox{!}{8.75cm}{\input{Figures/doe_variables_map_v2.tikz}}
    \caption{\footnotesize {Assignment of the main decision variables on a radial feeder: import at the slack, uncontrolled load, connection-point DOE on net DER, and behind-the-meter Wind, PV and BESS.}}
    \label{fig:variable_map}
\end{figure}
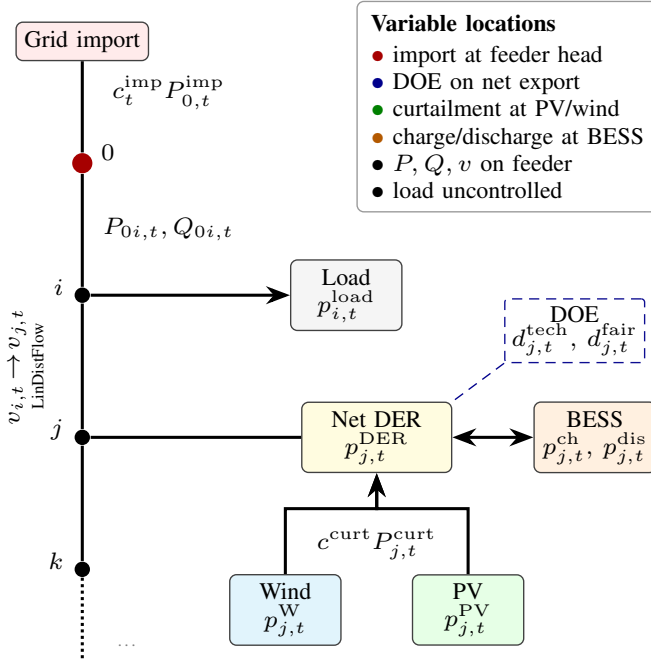

Throughout this section, the formulation will be discussed, based on individual components for it. Fig. \ref{fig:variable_map} showcases a simplified view for the main elements referenced in the following equations.
Furthermore, let $\mathcal{N}$ and $\mathcal{E}$ denote the sets of buses and branches of the radial feeder, $\mathcal{T}$ the set of time periods indexed by $t$, $\mathcal{P}$ the set of prosumers indexed by $i$, $\mathcal{B}$ the set of BESS units indexed by $b$, and $\Omega$ the set of scenarios indexed by $\omega$. Bus~$0$ denotes the substation (slack) bus.

\subsection{Objective Function}\label{sec:objective}
The operational objective combines import cost, renewable curtailment, BESS cycling and fairness-related penalties. \textcolor{black}{Conceptually, the formulation of the resulting multi-period problem can be described in the following compact form:}
\begin{align}
\underset{P_{0,t}^{\text{imp}}, P_{i,t}^{\text{curt}}, P_{b,t}^{\text{ch}}, P_{b,t}^{\text{dis}}} {\mbox{minimize}} \quad & \sum_{t \in \mathcal{T}} \Big(
c_t^{\text{imp}} P_{0,t}^{\text{imp}} 
+ c^{\text{curt}} \sum_{i \in \mathcal{P}} P_{i,t}^{\text{curt}} \nonumber \\
&+ c^{\text{BESS}} \sum_{b \in \mathcal{B}} (P_{b,t}^{\text{ch}} + P_{b,t}^{\text{dis}})
\Big) 
+ c^{\text{fair}} \Gamma
\label{eq:objective_align}
\end{align}
where $P_{0,t}^{\text{imp}}$ is the active power imported at the substation at time~$t$, priced at the import tariff $c_t^{\text{imp}}$; $P_{i,t}^{\text{curt}}$ is the renewable curtailment of prosumer~$i$ at time~$t$, penalized at unit cost $c^{\text{curt}}$; $P_{b,t}^{\text{ch}}$ and $P_{b,t}^{\text{dis}}$ are the charging and discharging powers of BESS unit~$b$, jointly penalized at cycling cost $c^{\text{BESS}}$; and $\Gamma$ is the maximum cumulative curtailment ratio from the fairness subproblem in Section~\ref{sec:doe}, with weight $c^{\text{fair}}$.

The objective is not designed only to minimize cost. It also internalizes renewable energy curtailment and the cost of storage cycling, while the fairness term represents the social objective of avoiding systematic discrimination against electrically remote prosumers. This structure is important because the cheapest dispatch may be technically feasible but unfair, whereas the fairest dispatch may curtail excessive renewable energy if no efficiency budget is imposed.

\vspace{-2mm}
\subsection{Network and Branch Flow Model}

For each branch $(i,j) \in \mathcal{E}$ and time $t \in \mathcal{T}$, let $P_{ij,t}$ and $Q_{ij,t}$ denote the active and reactive power flows from bus~$i$ to bus~$j$, and let $v_{i,t}$ denote the squared voltage magnitude at bus~$i$. The LinDistFlow \textcolor{black}{(linear distribution flow, a linear approximation tailored towards radial distribution networks) equations)} used inside the DOPF\textcolor{black}{, which simplifies the nonlinear power flow equations,} follows the branch-flow modelling tradition for radial feeders \cite{baran1989,farivar2013}:
\begin{subequations}\label{eq:branch_flow_model}
\begin{align}
%
%
P_{ij,t} = p^{\text{load}}_{j,t} - p^{\text{DER}}_{j,t} 
+ p^{\text{ch}}_{j,t} - p^{\text{dis}}_{j,t} 
+ \sum_{k: j \rightarrow k} P_{jk,t},
\label{eq:branch_active_power} \\
%
Q_{ij,t} = q^{\text{load}}_{j,t} - q^{\text{DER}}_{j,t} 
+ \sum_{k: j \rightarrow k} Q_{jk,t}
\label{eq:branch_reactive_power}, \\
%
v_{j,t} = v_{i,t} - 2 \big( r_{ij} P_{ij,t} + x_{ij} Q_{ij,t} \big)
\label{eq:voltage_drop}
%
\end{align}
\end{subequations}
where $p^{\text{load}}_{j,t}$ and $q^{\text{load}}_{j,t}$ are the active and reactive demands at bus~$j$; $p^{\text{DER}}_{j,t}$ and $q^{\text{DER}}_{j,t}$ are the net active and reactive injections from DERs connected at bus~$j$; $p^{\text{ch}}_{j,t}$ and $p^{\text{dis}}_{j,t}$ are the BESS charging and discharging powers at that bus; the sum over $k:j\rightarrow k$ aggregates flows into the children of bus~$j$ in the radial orientation; and $r_{ij}$ and $x_{ij}$ are the branch resistance and reactance.


The operating limits are
\begin{subequations} 
\begin{align}
\underline{V}^{2} \leq v_{i,t} \leq \overline{V}^{2}, \qquad \forall i \in \mathcal{N},\; t \in \mathcal{T},
\label{eq:voltage_limits}\\
P_{ij,t}^2 + Q_{ij,t}^2 \leq \overline{S}_{ij}^{2}, \qquad \forall (i,j) \in \mathcal{E},\; t \in \mathcal{T},
\label{eq:thermal_limits}
\end{align}
\end{subequations}
where $\underline{V}$ and $\overline{V}$ are the minimum and maximum admissible voltage magnitudes, and $\overline{S}_{ij}$ is the apparent-power (thermal) rating of branch $(i,j)$.

Although the DOPF uses the linearized branch-flow model to improve tractability in the distributed stage, the resulting dispatch is later checked through independent AC power flow. This validation step is necessary because linear power flow models are approximations and must be assessed against nonlinear AC feasibility when DER penetration is high \cite{bernstein2017,thurner2018,zamzam2018}.

\subsection{Technical and Fair Dynamic Operating Envelopes}\label{sec:doe}
The technical DOE maximizes the feasible export capacity of all prosumers while respecting network and DER constraints:
\begin{equation}
\max \sum_{t \in \mathcal{T}} \sum_{i \in \mathcal{P}} d_{i,t}^{\text{tech}}
\label{eq:tech_doe_obj}
\end{equation}
where $d_{i,t}^{\text{tech}}$ is the technical export envelope of prosumer~$i$ at time~$t$, i.e., the maximum net export at the connection point that remains network-feasible under the LinDistFlow model~\eqref{eq:branch_active_power}--\eqref{eq:thermal_limits}. 
The aggregate technical DOE is
\begin{equation}
D_{t}^{\text{tech}} = \sum_{i \in \mathcal{P}} d_{i,t}^{\text{tech}}
\label{eq:tech_doe_agg}
\end{equation}

A dynamic restriction factor $\beta_{t}\in[0,1]$ defines the aggregate export budget
\begin{equation}
D_{t}^{\text{cap}} = \beta_{t} D_{t}^{\text{tech}},
\label{eq:doe_budget}
\end{equation}
so that $\beta_{t}<1$ deliberately leaves unused hosting capacity relative to the purely technical allocation. 
The final DOE assigned to each prosumer is constrained by
\begin{equation}
0 \leq d_{i,t}^{\text{fair}} \leq d_{i,t}^{\text{tech}}, 
\quad \sum_{i \in \mathcal{P}} d_{i,t}^{\text{fair}} \leq D_{t}^{\text{cap}},
\label{eq:fair_doe_limits}
\end{equation}
where $d_{i,t}^{\text{fair}}$ is the fair (budgeted) export envelope at the same connection point.

Let the available and accepted renewable energies of prosumer~$i$ over the horizon be
\begin{equation}
E_{i}^{\text{av}} = \sum_{t \in \mathcal{T}} p_{i,t}^{\text{ren,av}} \, \Delta t,
\label{eq:E_av}
\end{equation}

\begin{equation}
E_{i}^{\text{acc}} = \sum_{t \in \mathcal{T}} p_{i,t}^{\text{acc}} \, \Delta t,
\label{eq:E_acc}
\end{equation}
where $p_{i,t}^{\text{ren,av}}$ is the available renewable generation, $p_{i,t}^{\text{acc}}$ is the accepted (uncurtailed) renewable injection, and $\Delta t$ is the duration of each time interval. \textcolor{black}{We can also write here that $p^{\text{acc}}_{j,t} - p^{\text{load}}_{j,t} \leq d_{i,t}$, which  determines that the accepted injection plus the local load consumption are less or equal to whichever envelope we are using.}
The cumulative curtailment ratio is
\begin{equation}
\chi_{i} = \frac{E_{i}^{\text{av}} - E_{i}^{\text{acc}}}{E_{i}^{\text{av}} + \varepsilon},
\label{eq:chi}
\end{equation}
where $\varepsilon>0$ is a small constant that avoids division by zero when $E_{i}^{\text{av}}=0$. 

The fairness subproblem minimizes the largest cumulative curtailment ratio:
\begin{equation}
\min \; \Gamma \quad \text{s.t.} \quad \chi_{i} \leq \Gamma, \quad \forall i \in \mathcal{P}
\label{eq:fairness_minmax}
\end{equation}
\textcolor{black}{selecting} the resulting maximum ratio $\Gamma$ \textcolor{black}{across all prosumers $i$, which then} enters the operational objective~\eqref{eq:objective_align}.

To preserve energy efficiency, fairness is constrained by an admissible curtailment budget \textcolor{black}{$E^{\text{adm}}$}:
\begin{equation}
E^{\text{curt}} \leq E^{\text{curt, tech}} + \Delta E^{\text{adm}} 
= E^{\text{curt, tech}} + \delta E^{\text{ren, av}},
\label{eq:curt_budget}
\end{equation}
where $E^{\text{curt}}$ is the total renewable curtailment under the fair allocation, $E^{\text{curt, tech}}$ is the curtailment of the technical-DOE benchmark, $E^{\text{ren, av}}$ is the aggregate available renewable energy, and $\delta\geq 0$ is the admissible fractional increase in curtailment relative to that benchmark.

\subsection{BESS and Scenario Recourse}

The BESS model couples decisions across time through the state-of-charge dynamics. For each unit $b\in\mathcal{B}$ and time $t\in\mathcal{T}$,
\begin{subequations} \label{eq:bess_model_complete}
\begin{align}
\mathrm{SOC}_{b,t+1} = \mathrm{SOC}_{b,t} + \eta_{c} P^{ch}_{b,t} \Delta t 
- \frac{1}{\eta_{d}} P^{dis}_{b,t} \Delta t,
\label{eq:soc}\\
\underline{\mathrm{SOC}}_{b} \leq \mathrm{SOC}_{b,t} \leq \overline{\mathrm{SOC}}_{b},
\label{eq:soc_limits}\\
0 \leq P_{b,t}^{ch} \leq \overline{P}_b^{ch},
\label{eq:pch_limits}\\
0 \leq P_{b,t}^{dis} \leq \overline{P}_b^{dis},
\label{eq:pdis_limits}
\end{align}
\end{subequations}
where $\mathrm{SOC}_{b,t}$ is the state of charge, $\eta_{c},\eta_{d}\in(0,1]$ are the charging and discharging efficiencies, and $\overline{P}_b^{ch}$, $\overline{P}_b^{dis}$ are the maximum charging and discharging rates.

For each scenario $\omega\in\Omega$, demand and renewable availability are scaled relative to the nominal profiles as
\begin{subequations} \label{eq:scenario)detail_model}
\begin{align}
p_{i,t}^{\mathrm{load},\omega} = \alpha_{d}^{\omega} p_{i,t}^{\mathrm{load}},
\label{eq:load_scenario}\\
P_{i,t}^{\mathrm{PV},\omega} = \alpha_{\mathrm{PV}}^{\omega} P_{i,t}^{\mathrm{PV}},
\label{eq:pv_scenario}\\
P_{i,t}^{\mathrm{W},\omega} = \alpha_{\mathrm{W}}^{\omega} P_{i,t}^{\mathrm{W}},
\label{eq:wind_scenario}
\end{align}
\end{subequations}
where $p_{i,t}^{\mathrm{load}}$, $P_{i,t}^{\mathrm{PV}}$ and $P_{i,t}^{\mathrm{W}}$ are the nominal demand, PV and wind profiles, and $\alpha_{d}^{\omega}$, $\alpha_{\mathrm{PV}}^{\omega}$, $\alpha_{\mathrm{W}}^{\omega}$ are the corresponding scenario scaling factors. 

The fair DOEs remain first-stage (scenario-independent) decisions, 
\begin{equation}
d_{i,t}^{\mathrm{DOE},\omega} = d_{i,t}^{\text{fair}}, \qquad \forall i\in\mathcal{P},\; t\in\mathcal{T},\; \omega\in\Omega,
\label{eq:doe_first_stage}
\end{equation}
while BESS powers and states of charge are recourse variables $\{P_{b,t}^{ch,\omega},\, P_{b,t}^{dis,\omega},\, \mathrm{SOC}_{b,t}^{\omega}\}$ that adapt to each $\omega$ under~\eqref{eq:soc}--\eqref{eq:pdis_limits} and the fixed envelopes~\eqref{eq:doe_first_stage}.

\subsection{Distributed ADMM Coordination}\label{sec:admm}
The feeder is decomposed into electrical regions $r\in\mathcal{R}$. At each interface $e=(i,j)$ between neighbouring regions, each region holds a local copy of the interface active power, reactive power and squared voltages:
\begin{equation}
y_{r,e,t} = 
\begin{bmatrix}
P_{e,t}^{r} &
Q_{e,t}^{r} &
v_{i,t}^{r} &
v_{j,t}^{r}
\end{bmatrix}^{\!\top}.
\label{eq:admm_y}
\end{equation}

Let $x_r$ collect the local primal variables of region~$r$, and let $f_r(x_r)$ denote its local contribution to~\eqref{eq:objective_align} subject to the regional LinDistFlow and device constraints. Each regional subproblem at ADMM iteration~$k$ is
\begin{equation}
x_r^{k+1} = \arg\min_{x_r} \; f_r(x_r) + \frac{\rho^{k}}{2} \left\| y_r(x_r) - z^k + u_r^k \right\|_2^2,
\label{eq:admm_primal}
\end{equation}
where $\rho^{k}>0$ is the penalty parameter used at iteration $k$, $z$ is the consensus (global) interface variable, and $u_r$ is the scaled dual variable associated with region~$r$. 

The consensus and dual updates are
\begin{equation}
z_e^{k+1} = \frac{1}{|R_e|} \sum_{r \in R_e} \left( y_{r,e}^{k+1} + u_{r,e}^k \right),
\label{eq:admm_z}
\end{equation}
\begin{equation}
u_{r,e}^{k+1} = u_{r,e}^{k} + y_{r,e}^{k+1} - z_{e}^{k+1},
\label{eq:admm_u}
\end{equation}
where $R_e\subseteq\mathcal{R}$ is the set of regions that share interface~$e$. 
The primal and dual residuals are
\begin{equation}
r^{k+1} = \left\| y^{k+1} - z^{k+1} \right\|_2,
\label{eq:admm_r}
\end{equation}
\begin{equation}
s^{k+1} = \rho^{k} \left\| z^{k+1} - z^k \right\|_2.
\label{eq:admm_s}
\end{equation}

Convergence is declared when $r^{k} \leq \epsilon^{\mathrm{pri}}$ and $s^{k} \leq \epsilon^{\mathrm{dual}}$, following the standard residual test of ADMM \cite{boyd2011}. $\rho^{k}$ is updated as per \cite{gebbran2023}.

The final dispatch is then fixed and validated by AC power flow, using
%
\begin{align}
\Delta V^{\mathrm{AC}} &= \max_{i\in\mathcal{N},\,t\in\mathcal{T}} \left| V^{\mathrm{AC}}_{i,t} - V^{\mathrm{DOPF}}_{i,t} \right|, \label{eq:ac_dV} \\
\Delta P^{\mathrm{AC}}_{0,\mathrm{corr}} &= \max_{t\in\mathcal{T}} \left| P^{\mathrm{AC}}_{0,t} - \left( P^{\mathrm{DOPF}}_{0,t} + P^{\mathrm{AC}}_{\mathrm{loss},t} \right) \right|,
\label{eq:ac_dP}
\end{align}
where $V^{\mathrm{AC}}_{i,t}$ and $P^{\mathrm{AC}}_{0,t}$ are the AC power-flow voltage and substation import, $V^{\mathrm{DOPF}}_{i,t}$ and $P^{\mathrm{DOPF}}_{0,t}$ are the corresponding LinDistFlow/DOPF quantities, and $P^{\mathrm{AC}}_{\mathrm{loss},t}$ is the AC network loss used to correct the substation power balance.

\subsection{Fairness and Validation Indicators}

After solving each scenario, fairness is quantified using the renewable acceptance ratio: 
\begin{equation}
\eta_i = \frac{E_i^{\text{acc}}}{E_i^{\text{av}} + \varepsilon},
\label{eq:eta}
\end{equation}
with $E_i^{\text{acc}}$ and $E_i^{\text{av}}$ as in~\eqref{eq:E_acc}--\eqref{eq:E_av}. The Jain fairness index is
\begin{equation}
\mathrm{JFI} = \frac{\left( \sum_{i \in \mathcal{P}} \eta_i \right)^2}{|\mathcal{P}| \sum_{i \in \mathcal{P}} \eta_i^2},
\label{eq:jfi}
\end{equation}
which equals one under perfect equality of acceptance ratios and decreases as disparity grows \cite{jain1984}. Curtailment inequality is also measured by the Gini coefficient
\begin{equation}
\mathrm{Gini} = \frac{\sum_{i \in \mathcal{P}} \sum_{\ell \in \mathcal{P}} \left| x_i - x_{\ell} \right|}{2 |\mathcal{P}| \sum_{i \in \mathcal{P}} x_i},
\label{eq:gini}
\end{equation}
where $x_i := E_i^{\text{av}} - E_i^{\text{acc}}$ is the cumulative renewable curtailment energy of prosumer~$i$ over the horizon. A Gini coefficient of zero indicates identical curtailment across prosumers, while larger values indicate stronger inequality.



All simulations were executed on a computer equipped with \textcolor{black}{an 11th Gen Intel(R) Core(TM) i5-1135G7 @ (2.42 GHz processor)}, and 8,00 GB of RAM, running 64-bit operating system, x64-based processor. This computational setup was used to obtain all numerical results, including the multi-period OPF solution, DOE allocation, fairness indicators, ADMM convergence curves, and AC validation results.

\section{Proposed Algorithm}

Algorithms~\ref{algorithm:DOE-fair} and~\ref{algorithm:scenario-admm} summarize the proposed workflow. Algorithm~\ref{algorithm:DOE-fair} first builds the radial model, computes technical export envelopes, and solves the budgeted fairness allocation. Algorithm~\ref{algorithm:scenario-admm} then, for each scenario, keeps the fair DOEs fixed as first-stage decisions, coordinates regional DOPF subproblems by ADMM, and validates the converged dispatch with independent AC power flow. The two-stage pipeline emphasizes that technical limits, fairness, storage recourse, distributed coordination and physical validation are coupled parts of one method. \textcolor{black}{Note that \eqref{eq:objective_align} is not implemented directly, but instead, the fair envelopes are calculated in the initial stage prior to the operational DOPF, broken down in this two-stage framework.}

\begin{figure}[!t]
\begin{algorithm}[H]
  \caption{Establishment of the DOE-Fair DOPF Model}
  \label{algorithm:DOE-fair}
  \begin{algorithmic}[1]
    \STATE \textbf{Data inputs:} network topology and line data; demand/PV/wind profiles; BESS parameters; scenario set $\Omega$; voltage/thermal limits; budget factor $\beta_t$; admissible curtailment parameter $\delta$; interval length $\Delta t$.
    \STATE \textbf{Initialization:} define $\mathcal{T}$, $\mathcal{P}$, $\mathcal{B}$ and $\Omega$; build the radial feeder $G(\mathcal{N},\mathcal{E})$; partition into regions $\mathcal{R}$; identify interface branches and consensus variables $y_{r,e,t}$ as in~\eqref{eq:admm_y}.
    \STATE Solve the technical multi-period DOPF~\eqref{eq:tech_doe_obj} subject to~\eqref{eq:branch_active_power}--\eqref{eq:thermal_limits} and device limits~\eqref{eq:soc}--\eqref{eq:pdis_limits}.
    \STATE Recover individual technical DOEs $d_{i,t}^{\mathrm{tech}}$ for all $i\in\mathcal{P}$, $t\in\mathcal{T}$.
    \STATE Compute the aggregate technical DOE~\eqref{eq:tech_doe_agg}.
    \STATE Apply the dynamic export budget~\eqref{eq:doe_budget}.
    \STATE Solve the fairness allocation~\eqref{eq:fairness_minmax} subject to~\eqref{eq:fair_doe_limits} and~\eqref{eq:curt_budget}, using~\eqref{eq:E_av}--\eqref{eq:chi}.
    \STATE \textbf{Outputs:} fair DOEs $d_{i,t}^{\mathrm{fair}}$, regional partition $\mathcal{R}$, and interface variables for Algorithm~\ref{algorithm:scenario-admm}.
  \end{algorithmic}
\end{algorithm}
\end{figure}

\begin{figure}[!t]
\begin{algorithm}[H]
  \caption{Scenario-Based Distributed DOPF via ADMM and AC Validation}
  \label{algorithm:scenario-admm}
  \begin{algorithmic}[1]
    \STATE \textbf{Data inputs:} $G(\mathcal{N},\mathcal{E})$; regional partition $\mathcal{R}$; fair DOEs $d_{i,t}^{\mathrm{fair}}$ from Algorithm~\ref{algorithm:DOE-fair}; scenario set $\Omega$; BESS data; ADMM parameters $\rho_0$, $\varepsilon^{\mathrm{pri}}$, $\varepsilon^{\mathrm{dual}}$.
    \FOR{each scenario $\omega\in\Omega$}
        \STATE Scale demand, PV and wind via~\eqref{eq:load_scenario}--\eqref{eq:wind_scenario}.
        \STATE Fix first-stage DOEs~\eqref{eq:doe_first_stage}; treat $\{P_{b,t}^{ch,\omega},P_{b,t}^{dis,\omega},\mathrm{SOC}_{b,t}^{\omega}\}$ as recourse under~\eqref{eq:soc}--\eqref{eq:pdis_limits}.
        \STATE Initialize ADMM: $k\leftarrow 0$, $\rho^{k}\leftarrow\rho_0$, $z^{0}$, $u^{0}$.
        \WHILE{$r^{k}>\varepsilon^{\mathrm{pri}}$ \textbf{or} $s^{k}>\varepsilon^{\mathrm{dual}}$}
            \FOR{each region $r\in\mathcal{R}$ \textbf{in parallel}}
                \STATE Given $z^{k}$ and $u^{k}$, solve the regional subproblem~\eqref{eq:admm_primal} subject to~\eqref{eq:branch_active_power}--\eqref{eq:thermal_limits} and~\eqref{eq:objective_align}.
                \STATE Return local interface copies $y_{r}^{k+1}$ as in~\eqref{eq:admm_y}.
            \ENDFOR
            \STATE Update consensus variables via~\eqref{eq:admm_z}.
            \STATE Update dual variables via~\eqref{eq:admm_u}.
            \STATE Compute residuals~\eqref{eq:admm_r}--\eqref{eq:admm_s}; update $\rho^{k}$ as per  \cite{gebbran2023}; $k\leftarrow k+1$.
        \ENDWHILE
        \STATE Fix the converged DER/BESS dispatch for scenario~$\omega$.
        \STATE Run independent AC power flow for every $t\in\mathcal{T}$ and evaluate~\eqref{eq:ac_dV}--\eqref{eq:ac_dP}.
        \STATE Compute curtailment, cost, $\eta_i$~\eqref{eq:eta}, JFI~\eqref{eq:jfi}, Gini~\eqref{eq:gini}, and ADMM metrics.
    \ENDFOR
    \STATE \textbf{Outputs:} dispatch schedules, curtailment, costs, fairness indices, ADMM convergence and AC validation results.
  \end{algorithmic}
\end{algorithm}
\end{figure}


\section{Case Study}
The methodology is evaluated on the IEEE 33-bus radial feeder $G(\mathcal{N},\mathcal{E})$ \cite{baran1989}, with $|\mathcal{N}|=33$ buses and a single slack at bus~$0$. As illustrated in Fig.~\ref{fig:variable_map}, each prosumer connection can host net DER injection $p^{\mathrm{DER}}_{j,t}$ subject to a connection-point DOE $d_{j,t}^{\mathrm{tech}}$/$d_{j,t}^{\mathrm{fair}}$, while uncontrolled demand $p^{\mathrm{load}}_{i,t}$ is treated separately from behind-the-meter Wind, PV and BESS. The feeder is partitioned into nine electrical regions $r\in\mathcal{R}$ for the ADMM stage in Algorithm~\ref{algorithm:scenario-admm}, with interface consensus variables~\eqref{eq:admm_y}.

The operating horizon is $\mathcal{T}=\{1,\ldots,24\}$ with $\Delta t=\SI{1}{\hour}$. Fifteen PV units and eight wind units supply the renewable profiles $P_{i,t}^{\mathrm{PV}}$ and $P_{i,t}^{\mathrm{W}}$, while eight BESS units $b\in\mathcal{B}$ provide temporal recourse through~\eqref{eq:soc}--\eqref{eq:pdis_limits}. The import tariff $c_t^{\mathrm{imp}}$ enters the objective~\eqref{eq:objective_align}. Unless otherwise stated, the dynamic DOE budget uses $\beta_t=0.70$ over the midday window $t\in\{10,\ldots,16\}$ in~\eqref{eq:doe_budget}, and every converged LinDistFlow dispatch is checked by independent AC power flow via~\eqref{eq:ac_dV}--\eqref{eq:ac_dP}.

\begin{figure}[!t]
    \centering
    \includegraphics[width=\columnwidth]{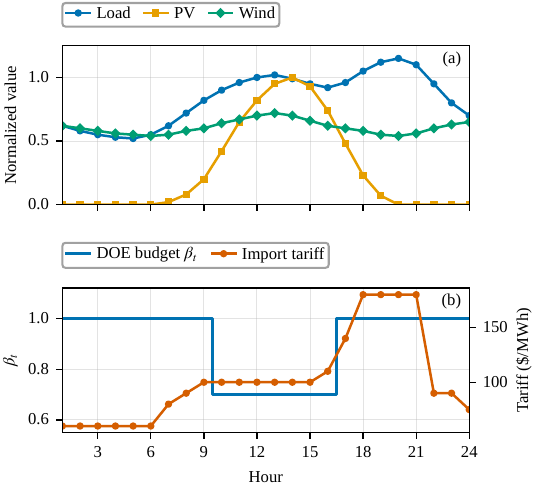}
    \caption{Input data used in the 24-hour case study: (a)~normalized load, PV and wind profiles; (b)~dynamic DOE budget factor~$\beta_t$ (left axis) and import tariff in \$/MWh (right axis).}
    \label{fig:input_profiles_doe_tariff}
\end{figure}

Fig.~\ref{fig:input_profiles_doe_tariff} shows the normalized load, PV and wind profiles, the DOE budget factor $\beta_t$, and the import tariff $c_t^{\mathrm{imp}}$. The restrictive DOE window ($\beta_t=0.70$ from hour~10 to~16) coincides with peak PV availability and is therefore the most critical period for renewable curtailment $P_{i,t}^{\mathrm{curt}}$ and fair envelope allocation. Outside that window, $\beta_t=1$ recovers the full technical aggregate export $D_t^{\mathrm{tech}}$.

Table~\ref{tab:Case_Study_Scenarios} lists the simulated cases. They are organized incrementally: S1--S3 establish network and DER baselines; S4 compares technical versus budgeted DOEs; S5 activates fairness~\eqref{eq:fairness_minmax} under alternative allocation rules; S6 introduces scenario scalings~\eqref{eq:load_scenario}--\eqref{eq:wind_scenario} with fixed first-stage envelopes~\eqref{eq:doe_first_stage}; and S7 reports the AC validation layer.



\begin{table}[!t]
\centering
\caption{Case-study scenarios}
\label{tab:Case_Study_Scenarios}
\begin{tabular}{llp{0.42\columnwidth}}
\hline
\textbf{Code} & \textbf{Configuration} & \textbf{Purpose} \\
\hline
S1 & No DER & Reference losses and voltages \\
S2 & PV + wind & Renewables without BESS \\
S3 & PV + wind + BESS & Temporal coupling via $\mathrm{SOC}_{b,t}$ \\
S4-T & Technical DOE & Maximize $d_{i,t}^{\mathrm{tech}}$ in~\eqref{eq:tech_doe_obj} \\
S4-R & Restrictive DOE & Budget~\eqref{eq:doe_budget} with $\beta_t=0.70$ (10--16\,h) \\
S5: E/P/U/TP & Fair DOE & Equal, proportional, uniform and temporal allocations under~\eqref{eq:fair_doe_limits}--\eqref{eq:curt_budget} \\
S6: $\omega_1$--$\omega_6$ & Uncertainty & Scenario scalings $\alpha_d^{\omega}$, $\alpha_{\mathrm{PV}}^{\omega}$, $\alpha_{\mathrm{W}}^{\omega}$ \\
S7 & AC validation & Metrics~\eqref{eq:ac_dV}--\eqref{eq:ac_dP} on the converged dispatch \\
\hline
\end{tabular}
\end{table}


\section{Results and Discussion}

All optimization models are implemented in Julia using JuMP~\cite{Lubin2023} and solved with HiGHS~\cite{Huangfu2018}, on an Intel Core~i5-1135G7 (2.42~GHz) laptop with 8~GB of RAM. The 24-h LinDistFlow DOPF (Algorithms~\ref{algorithm:DOE-fair}--\ref{algorithm:scenario-admm}, including the six operating conditions) required a median wall-clock time of approximately 23~min~47~s over five timed runs after one warm-up, excluding data loading, plotting and file export. Technical and fair envelopes are computed first, then each operating condition is coordinated by regional ADMM and checked by AC power flow. Without this separation (i.e., under a purely technical OPF/DOE or a single-interval fairness objective) the equity-efficiency trade-off, the cumulative curtailment limit $\Gamma$, and the scenario-dependent BESS recourse would remain entangled and difficult to quantify.

\textcolor{black}{Unless stated otherwise, the quantitative results below for ``operating conditions'' $\omega_1$--$\omega_6$ refer to the six uncertainty cases of Table~\ref{tab:Case_Study_Scenarios} (case~S6), which were chosen as cases of interest with critical conditions. Figs.~\ref{fig:admm_convergence_residuals}--\ref{fig:cost_iterations_scenarios} label these same six conditions as S1--S6 (nominal, high PV/low load, high wind, peak/low RES, night peak, and mixed congestion).}


\subsection{Technical and Fair DOE Allocation}

Applying Algorithm~\ref{algorithm:DOE-fair}, the technical stage~\eqref{eq:tech_doe_obj} yields $59.6778$~MWh of renewable availability, \textcolor{black}{$2.0898$}~MWh of technical curtailment and \textcolor{black}{$28.7860$}~MWh of aggregate technical DOE $\sum_{t}D_{t}^{\mathrm{tech}}$ over $\mathcal{T}$. The fairness stage~\eqref{eq:fairness_minmax} returns \textcolor{black}{$\Gamma=0.113095$}, i.e., a maximum cumulative curtailment ratio of \textcolor{black}{$11.31\%$}, with total fair curtailment \textcolor{black}{$5.7741$}~MWh. The admissible curtailment budget~\eqref{eq:curt_budget} is \textcolor{black}{$19.9931$}~MWh, so the fair solution remains inside the efficiency allowance $\delta$.

\begin{figure}[!t]
        \centering
        \includegraphics[width=\columnwidth]{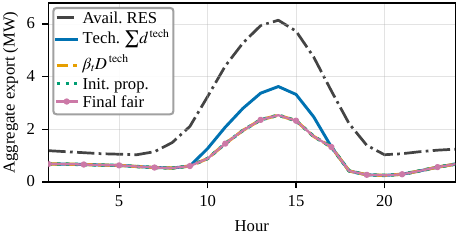}
        \caption{Aggregate renewable availability, technical DOE $\sum_{i}d_{i,t}^{\mathrm{tech}}$, dynamic budget $\beta_t D_{t}^{\mathrm{tech}}$, initial proportional allocation and final fair DOE over $24$~h.}
        \label{fig:aggregate_doe}
\end{figure}

Fig.~\ref{fig:aggregate_doe} shows the corresponding aggregate trajectories. Midday renewable availability (dash-dotted) substantially exceeds the technical aggregate DOE $\sum_{i}d_{i,t}^{\mathrm{tech}}$ (solid blue) — peaking near $6.1$~MW versus about $3.6$~MW of technical export capacity at hour~$14$ — confirming that network limits, not resource scarcity, drive curtailment. During the restrictive window of Fig.~\ref{fig:input_profiles_doe_tariff}, the budget $\beta_t D_{t}^{\mathrm{tech}}$~\eqref{eq:doe_budget}, the initial proportional allocation and the final fair allocation $\sum_{i}d_{i,t}^{\mathrm{fair}}$ coincide near $2.5$~MW and lie about $1.1$~MW below the technical DOE. Outside that window ($\beta_t=1$), fair and technical aggregates collapse onto each other. This is precisely the mechanism that a static export limit or a pure technical DOE cannot expose: fairness is enforced only when the budget binds, and the cost of fairness appears as additional midday curtailment relative to~\eqref{eq:tech_doe_obj}.

\begin{figure}[!t]
        \centering
        \includegraphics[width=\columnwidth]{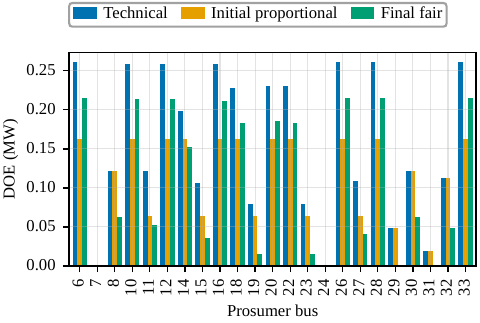}
        \caption{DOE allocation at hour~14 for the prosumer buses: technical DOE, initial proportional allocation and final fair allocation.}
       \label{fig:doe_allocation_hour14}
\end{figure}

Fig.~\ref{fig:doe_allocation_hour14} zooms into hour~$14$, when PV peaks and $\beta_t=0.70$. Although many buses share a similar technical ceiling near $0.26$~MW, the final fair envelopes $d_{i,14}^{\mathrm{fair}}$ differ sharply across prosumers (e.g., comparatively generous allocations near buses~$12$, $16$, $26$, $28$ and $33$, versus strongly reduced envelopes at buses~$8$, $19$, $23$ and $27$, and near-zero fair export at buses~$29$ and~$31$). A purely technical allocation would privilege electrically favourable locations; the cumulative criterion~\eqref{eq:chi}--\eqref{eq:fairness_minmax} redistributes capacity under~\eqref{eq:fair_doe_limits}, which cannot be recovered from a snapshot OPF that ignores prior curtailment history.

\subsection{Nominal Dispatch and Voltage Performance}

\begin{figure}[!t]
    \centering
    \includegraphics[width=\columnwidth]{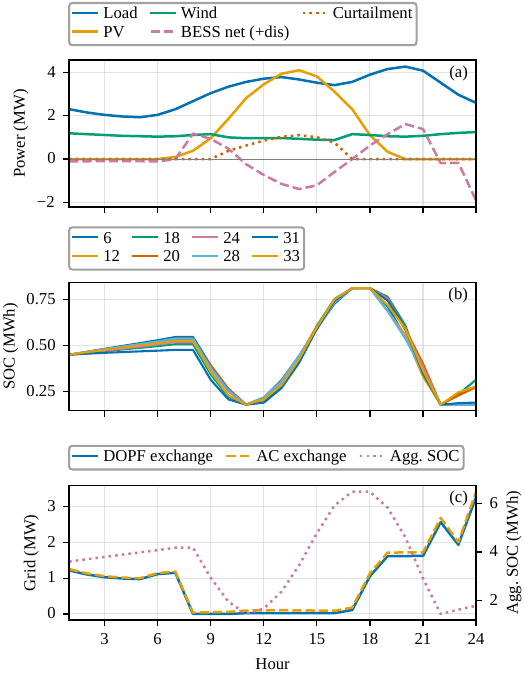}
    \caption{Nominal 24-hour operation: (a)~demand, renewable generation and BESS dispatch; (b)~individual BESS state of charge; (c)~substation exchange with aggregate BESS SOC.}
    \label{fig:nominal_dispatch_bess_grid}
\end{figure}

Under the nominal operating condition ($\omega_1$), Algorithm~\ref{algorithm:scenario-admm} coordinates renewable dispatch, BESS recourse and substation exchange subject to the fixed fair DOEs~\eqref{eq:doe_first_stage}. Fig.~\ref{fig:nominal_dispatch_bess_grid}(a) shows that midday PV dispatch rises sharply while wind remains comparatively flat; curtailment appears between hours~$9$ and~$16$, peaking near $1.1$~MW at hour~$14$ where the fair aggregate DOE binds (cf.\ Fig.~\ref{fig:aggregate_doe}). Net BESS power is briefly positive in the early morning, strongly negative (charging, down to about $-1.4$~MW) through the high-PV window, and positive again around the evening load peak near hour~$20$.

The individual and aggregate SOC trajectories in Fig.~\ref{fig:nominal_dispatch_bess_grid}(b)--(c) make the multi-period coupling~\eqref{eq:soc} explicit: all eight units follow a synchronized pattern — discharging toward a lower SOC by mid-morning, filling through midday to a common peak near hour~$17$, then depleting through the evening. Substation import drops to essentially zero while PV and charged storage cover demand (hours~$9$--$17$), then rises again as PV disappears and SOC is drawn down; the DOPF and AC exchange traces in the bottom panel already track closely, foreshadowing the dedicated validation in Fig.~\ref{fig:nominal_ac_validation}. A single-period DOE/OPF without SOC dynamics could not schedule this temporal shift; likewise, fixing BESS ahead of uncertainty would forfeit the recourse role used in~\eqref{eq:doe_first_stage}--\eqref{eq:pdis_limits}.

\begin{figure}[!t]
        \centering
        \includegraphics[width=\columnwidth]{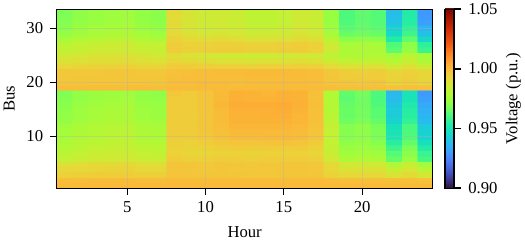}
        \caption{Nominal voltage magnitude heatmap (bus vs.\ hour) from the final $\omega_1$ dispatch.}
       \label{fig:nominal_voltage}
\end{figure}

Fig.~\ref{fig:nominal_voltage} reports the resulting voltage heatmap. Midday voltages rise under reverse power flow, while remote buses (higher indices) experience the deepest evening depression. All values remain above the hard lower limit $0.90$~p.u.\ adopted in~\eqref{eq:voltage_limits}, although several remote buses fall below $0.95$~p.u.\ after hour~$21$. Stricter statutory bands would therefore require additional reactive support or more conservative envelopes — again underscoring why DOE limits and multi-period storage must be co-designed rather than post-processed independently.

\subsection{\textcolor{black}{ADMM Convergence and Analysis for} Conditions $\omega_1$--$\omega_6$}

\begin{figure}[!t]
        \centering
        \includegraphics[width=\columnwidth]{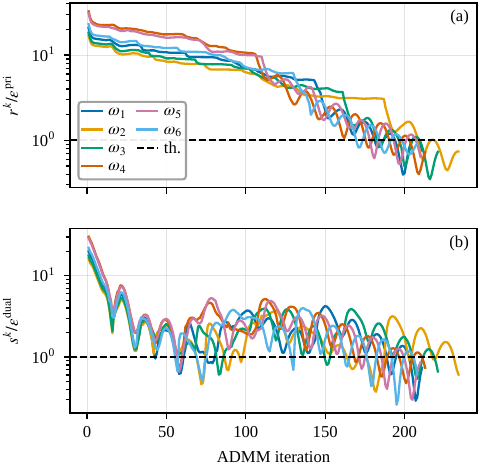}
        \caption{ADMM convergence for $\omega_1$--$\omega_6$: (a)~normalized primal residual and (b)~normalized dual residual. The dashed horizontal line denotes the normalized convergence threshold.}
        \label{fig:admm_convergence_residuals}
\end{figure}

Convergence of Algorithm~\ref{algorithm:scenario-admm} is monitored through the normalized residuals $r^{k}/\varepsilon^{\mathrm{pri}}$ and $s^{k}/\varepsilon^{\mathrm{dual}}$ associated with~\eqref{eq:admm_r}--\eqref{eq:admm_s}. Fig.~\ref{fig:admm_convergence_residuals} shows both residuals for all six operating conditions. The primal residual decays largely monotonically as regions agree on interface $(P,Q,v)$ copies~\eqref{eq:admm_y}. The dual residual is more oscillatory — as expected with multi-period BESS coupling and binding DOE limits — yet eventually remains below the unit threshold.

All six conditions converge within $208$--$234$ iterations (Table~\ref{tab:s6_metrics}): $\omega_1$ (nominal) in $211$ iterations, and $\omega_2$ (high PV/low load) in $234$ iterations, the largest count. The latter stresses export budgets, curtailment and interface flows simultaneously, consistent with prior experience that stressed DER coordination problems require more ADMM iterations~\cite{Gebbran_SGES}. Importantly, none of the conditions fails to meet both tolerances, which supports the calibrated regional decomposition for repeated DOE-constrained DOPF solves.

\begin{figure}[!t]
    \centering
    \includegraphics[width=\columnwidth]{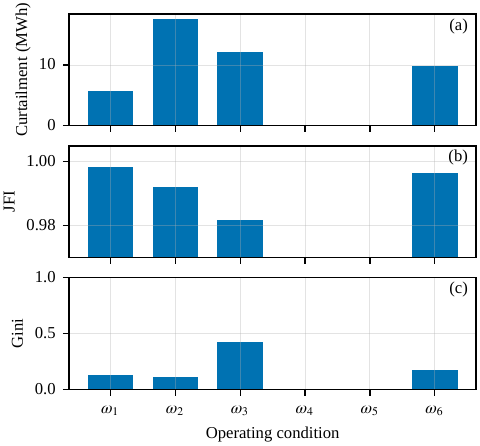}
    \caption{Fairness and curtailment indicators by operating condition: (a)~renewable curtailment, (b)~Jain fairness index, and (c)~Gini coefficient for $\omega_1$--$\omega_6$.}
    \label{fig:fairness_curtailment_metrics}
\end{figure}

\begin{figure}[!t]
    \centering
    \includegraphics[width=\columnwidth]{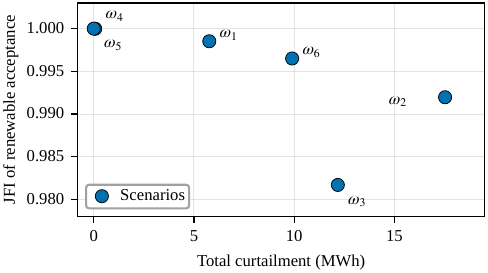}
    \caption{Efficiency--fairness map: total curtailment versus JFI for $\omega_1$--$\omega_6$.}
    \label{fig:eff_fair_map}
\end{figure}

\begin{figure}[!t]
    \centering
    \includegraphics[width=\columnwidth]{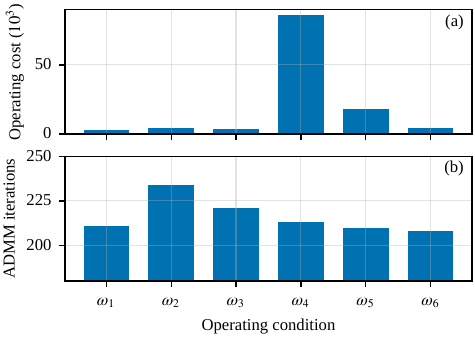}
    \caption{\textcolor{black}{Detailed solution indicators by operating condition: (a)~total operating cost, and (b)~total ADMM iterations.}}
    \label{fig:cost_iterations_scenarios}
\end{figure}

Table~\ref{tab:s6_metrics} and Figs.~\ref{fig:fairness_curtailment_metrics}--\ref{fig:cost_iterations_scenarios} summarize performance across $\omega_1$--$\omega_6$. The highest renewable curtailment occurs under $\omega_2$ (high PV/low load): $17.528$~MWh ($25.47\%$), followed by $\omega_3$ (high wind, $12.177$~MWh) and $\omega_6$ (mixed congestion, $9.898$~MWh). These are exactly the conditions in which connection-point envelopes~\eqref{eq:fair_doe_limits} must bind if feeder integrity is to be preserved — a static $5$~kW-style limit would either over-curtail in mild hours or violate voltages in these stressed hours.

Ex-post fairness remains strong wherever curtailment is material: JFI~\eqref{eq:jfi} stays above $0.98$, with $\omega_1$ and $\omega_6$ near $0.998$ (Fig.~\ref{fig:fairness_curtailment_metrics}). The Gini coefficient~\eqref{eq:gini} on curtailment energy $x_i$ is low for $\omega_1$, $\omega_2$ and $\omega_6$ ($0.11$--$0.17$) but rises under $\omega_3$ ($0.4216$), indicating that high-wind congestion still concentrates some curtailment on electrically weaker buses even when JFI is high. For $\omega_4$ and $\omega_5$ (peak/low RES and night peak), curtailment is negligible, \textcolor{black}{so JFI and Gini are omitted from Table~\ref{tab:s6_metrics} and from panels (b)--(c) of Fig.~\ref{fig:fairness_curtailment_metrics}; a near-zero curtailment total makes those indices numerically unstable. These two conditions instead require limited emergency demand response ($0.0807$ and $0.0137$~MWh) and are marked as not strictly feasible without that recourse, which in turn is an evidence that envelopes alone cannot replace flexibility when renewables are scarce.}

Fig.~\ref{fig:eff_fair_map} maps the efficiency--fairness trade-off in the $(E^{\mathrm{curt}},\mathrm{JFI})$ plane: $\omega_4$/$\omega_5$ sit at near-zero curtailment and unit JFI; $\omega_2$ is the efficiency extreme (largest curtailment) while still keeping $\mathrm{JFI}\approx0.992$; $\omega_3$ is the fairness outlier among high-curtailment cases. Fig.~\ref{fig:cost_iterations_scenarios} complements this view economically: operating cost explodes under $\omega_4$ (peak demand, low RES), with a secondary increase under $\omega_5$, whereas renewable-rich conditions keep cost low. ADMM iteration counts remain in a narrow band despite this cost disparity, reinforcing that computational effort is driven more by binding network/DOE structure than by the absolute cost level.

\begin{table}[!t]
\centering
\caption{Performance metrics for case~S6 operating conditions $\omega_1$--$\omega_6$}
\label{tab:s6_metrics}
\scriptsize
\setlength{\tabcolsep}{3pt}
\renewcommand{\arraystretch}{1.05}
\begin{tabular}{lccccccc}
\hline
\textbf{Condition} &
\textbf{Curt.} &
\textbf{Curt.} &
\textbf{JFI} &
\textbf{Gini} &
\textbf{Iter.} &
\textbf{Strict} &
\textbf{DR} \\
 &
\textbf{(MWh)} &
\textbf{(\%)} &
 &
 &
 &
 &
\textbf{(MWh)} \\
\hline
$\omega_1$ nominal & 5.763 & 9.66 & 0.9985 & 0.1244 & 211 & Yes & 0 \\
$\omega_2$ high PV/low load & 17.528 & 25.47 & 0.9920 & 0.1096 & 234 & Yes & 0 \\
$\omega_3$ high wind & 12.177 & 19.17 & 0.9817 & 0.4216 & 221 & Yes & 0 \\
$\omega_4$ peak/low RES & $\approx 0$ & $\approx 0$ & -- & -- & 213 & No & 0.0807 \\
$\omega_5$ night peak & $\approx 0$ & $\approx 0$ & -- & -- & 210 & No & 0.0137 \\
$\omega_6$ mixed congestion & 9.898 & 14.76 & 0.9965 & 0.1697 & 208 & Yes & 0 \\
\hline
\end{tabular}
\end{table}


\subsection{Independent AC Validation}

\begin{figure}[!t]
        \centering
        \includegraphics[width=\columnwidth]{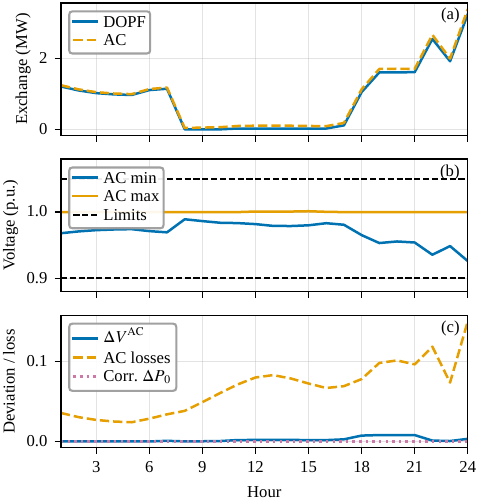}
        \caption{Nominal AC validation: (a)~substation exchange, (b)~voltage profile, and (c)~model deviations.} 
        \label{fig:nominal_ac_validation}
\end{figure}

Because Algorithm~\ref{algorithm:scenario-admm} optimizes a lossless LinDistFlow model~\eqref{eq:branch_active_power}--\eqref{eq:voltage_drop}, each converged schedule is re-evaluated with nonlinear AC power flow. Fig.~\ref{fig:nominal_ac_validation} reports the nominal ($\omega_1$) comparison. Substation exchange from DOPF and AC track closely, with AC slightly higher as expected from losses; minimum AC voltages remain above $0.90$~p.u.\ and maximum voltages near $1.00$~p.u., inside~\eqref{eq:voltage_limits}. Voltage deviation $\Delta V^{\mathrm{AC}}$~\eqref{eq:ac_dV} and corrected substation deviation $\Delta P_{0,\mathrm{corr}}^{\mathrm{AC}}$~\eqref{eq:ac_dP} stay near zero once AC losses are accounted for.

Across the evaluated conditions, $\max\Delta V^{\mathrm{AC}}$ remains below approximately $0.01$~p.u., no voltage or thermal violations appear under the $0.90$--$1.05$~p.u.\ band, and the corrected power metric stays near zero. Importantly, this AC step is an independent nonlinear power-flow check with the DER/BESS schedule held fixed, and not a separate AC OPF solve, so it does not provide a like-for-like optimizer runtime against the LinDistFlow DOPF reported above; a dedicated AC validation wall-clock was not measured in the submitted runs. The LinDistFlow timing nonetheless matters when envelopes must be refreshed repeatedly (e.g., intra-hour updates). The validation layer is therefore not cosmetic: without it, one could not claim that the distributed fair-DOE schedules remain physically meaningful under AC physics.

\subsection{Discussion of Technical Implications}

When renewable availability exceeds network export capability, the connection-point DOE decides which prosumers may inject and which are curtailed. Figs.~\ref{fig:aggregate_doe}--\ref{fig:doe_allocation_hour14} show that this decision is both technical and distributional: embedding fairness only inside a single OPF objective would obscure the explicit cost of moving from $2.11$~MWh to \textcolor{black}{$5.77$}~MWh of curtailment while enforcing \textcolor{black}{$\Gamma\leq 11.31\%$}. The cumulative criterion avoids myopic, location-driven curtailment that systematically penalizes remote buses over the day.

Treating BESS as scenario recourse, rather than as a first-stage schedule, is equally consequential. Fig.~\ref{fig:nominal_dispatch_bess_grid} shows storage shifting midday renewable energy toward the evening peak; Table~\ref{tab:s6_metrics} shows that even then, $\omega_4$ and $\omega_5$ still need a small\textcolor{black}{, residual demand response activation $p^{\text{DR}}_{j,t}$, which could be separately accounted for as a Value of Lost Load (VOLL) cost}. Those cases identify when additional flexibility or tighter envelopes are required — information that a single deterministic DOE computation would miss.

Finally, distributed ADMM coordination with AC validation addresses the practical barrier that motivates DOEs in the first place: direct OPF micro-management of every DER is undesirable and often infeasible~\cite{Chatzivasileiadis_2023_MicroFlex}, yet ignoring network physics is unsafe. The proposed pipeline publishes fair envelopes, solves regional DOPFs that respect them, and verifies AC feasibility — a combination that neither static limits nor standalone technical OPF provides. \textcolor{black}{Also, in practice, the DOPF allows for a parallel implementation, which reduces overall computation times, and therefore favors the implementation of the DOE calculation within a rolling horizon, which requires frequent re-computes.}

\vspace{-2mm}


\section{Conclusion}

This paper presents a two-stage multi-period framework for fair dynamic operating envelopes in active radial distribution networks. Technical connection-point envelopes are computed first from network-feasible DOPF; a budgeted cumulative fairness stage then redistributes export capacity under an admissible curtailment-efficiency limit, yielding an explicit equity-efficiency trade-off that single-objective OPF embeddings do not isolate. The resulting fair DOEs are held as first-stage decisions while BESS provides scenario-dependent recourse; the operational problem is solved by calibrated regional ADMM on a LinDistFlow model and validated by independent AC power flow (Algorithms~\ref{algorithm:DOE-fair}--\ref{algorithm:scenario-admm}).

On the IEEE~33-bus feeder, the technical stage curtails \textcolor{black}{$\SI{2.0898}{\mega\watt\hour}$}, whereas the fair allocation curtails \textcolor{black}{$\SI{5.7741}{\mega\watt\hour}$} and enforces a maximum cumulative curtailment ratio \textcolor{black}{$\Gamma=\SI{11.31}{\%}$}. Across six operating conditions, ADMM converges in $208$--$234$ iterations, Jain indices remain close to unity when curtailment is material, and AC validation reports no voltage/thermal violations under $0.90$--$1.05$~p.u.\ limits with $\Delta V^{\mathrm{AC}}$ below $0.01$~p.u. The analysis herein presented shows that co-designing multi-period fairness, which increases curtailment, and battery storage, which alleviates curtailment impact, offers a solid alternative to single-period DOE design, which would be otherwise more conservative — and such co-design requires an integrated multi-period approach, as proposed herein. 

Future work includes leveraging the distributed solution towards implementation in unbalanced three-phase feeders and scaling up to larger networks, applying the methodology towards probabilistic envelopes, experimenting alternative BESS participation layers in DOE generation, and creating rolling-horizon tests on real feeder data.

\bibliographystyle{ieeetr}
\bibliography{references}{}

\vskip -2\baselineskip plus -1fil
\begin{IEEEbiography}[{\includegraphics[width=1in,height=1.25in,clip,keepaspectratio]{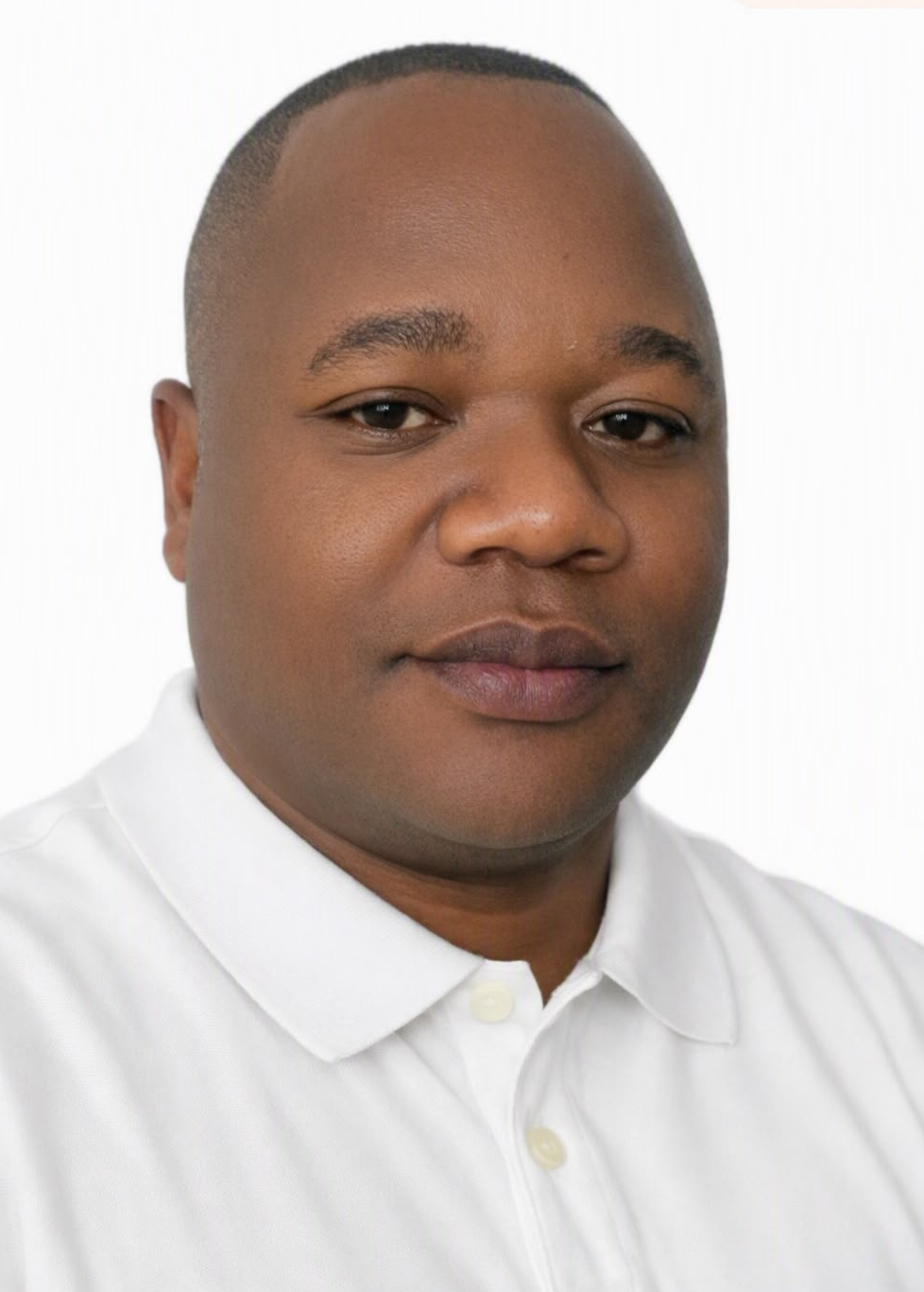}}]
{Pedro Salomão Quessongo} (S’25) received the B.Sc. degree in Electronic and Telecommunications Engineering from José Eduardo dos Santos University (UJES), Huambo, Angola, in 2013, He is currently pursuing a Master's degree in the Graduate Program in Electrical Engineering at the Federal University of Parana (UFPR), Curitiba, Brazil. From 2013 up to now, he has been working as an Electronics and telecommunications Engineer at the National Electricity Transmission Network (RNT-EP), Angola, and from 2018 up to date works as faculty member at the Higher Polytechnic Institute of Caála (ISPCAALA), Caála, Angola. His research interests include electric power systems, measurement, Automation and control system, protection of power system, as well as renewable energy and sustainability.
\end{IEEEbiography}

\vskip -2\baselineskip plus -1fil
\begin{IEEEbiography}
[{\includegraphics[width=1in,height=1.25in,clip,keepaspectratio]{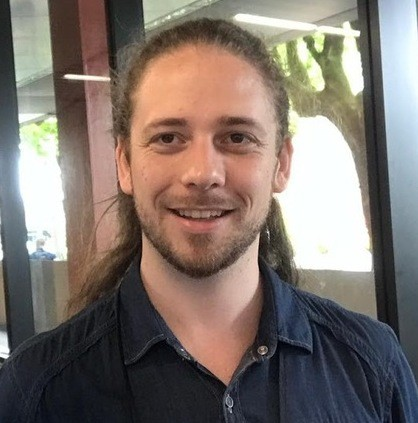}}]{Daniel Gebbran} (S'18-M'20) received the B.Sc. degree in electrical power engineering from the Federal University of Technology – Parana (UTFPR), Curitiba, Brazil, in 2014, the M.Sc. degree in electrical engineering and computer sciences from the University of California Irvine (UCI), USA, in 2017, and the Ph.D. degree from the University of Sydney, Australia, in 2021. He held a one-year postdoctoral researcher position at the Technical University of Denmark (DTU) through 2022, and currently is a postdoctoral researcher at the Federal University of Paraná since 2025, focusing on the development and implementation of machine learning and optimization algorithms in academic and industrial projects combining multiple DERs. Dr. Gebbran has worked in ERCOT, CAISO, as well as Brazilian, Australian and European energy markets over the past ten years with an emphasis on applied research, having worked on Bravos Energia and Equilibrium Energy previously. He is currently the Director of BESS Optimization in Caerus Commodities, overseeing development, implementation, and deployment of mathematical forecasting and optimization models for operation of over $\SI{4}{\giga\watt\hour} $  of large-scale batteries in US energy markets. 
\end{IEEEbiography}

\vskip -2\baselineskip plus -1fil
\begin{IEEEbiography}[{\includegraphics[width=1in,height=1.25in,clip,keepaspectratio]{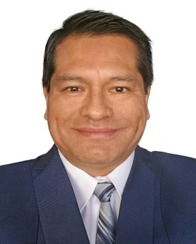}}]{Clodomiro Unsihuay-Vila} (S’01–M’09) received the degree in electrical engineering from the National University of the Center of Peru (UNCP), Peru, in 2000, the M.Sc. degree in electrical engineering from the Federal University of Maranhão, Brazil, in 2003, and the Ph.D. degree in electrical engineering from the Federal University of Itajubá, Brazil, in 2009. He completed the Sandwich-Doctoral Fellowship from the Pontifical University of Comillas, Madrid, Spain, from 2007 to 2008. He is currently a full Professor with the Electrical Engineering Department, Federal University of Paraná (UFPR), Brazil. He has more than 100 research publications in international journals and conference proceedings. His current research interests include modelling, analysis, planning, operation, control, optimization, integration of contemporary and future electrical power, energy systems, integration of disruptive energy technologies and the computational and IA and data-driven tools that enable secure, reliable, resilient, flexible, and sustainable power networks.

\end{IEEEbiography}

\color{red}

\end{document}

%% file: Figures/doe_variables_map_v2.tikz

\begin{tikzpicture}[
  x=1.05cm, y=1.05cm,
  >=Stealth,
  bus/.style={circle, fill=black, inner sep=1.6pt},
  slack/.style={circle, fill=red!65!black, inner sep=2.1pt},
  feeder/.style={line width=0.9pt},
  derbox/.style={
    draw=black!75, rounded corners=2pt, align=center,
    font=\scriptsize, inner sep=2.5pt, minimum width=1.15cm,
    fill=white
  },
  callout/.style={
    font=\scriptsize, align=center, fill=white, opacity=0.95,
    text opacity=1, inner sep=2pt, rounded corners=1pt
  },
]

\node[derbox, fill=red!8] (grid) at (0,3.6) {Grid import};
\node[callout] at (0.85,3.1) {$c_t^{\mathrm{imp}} P_{0,t}^{\mathrm{imp}}$};

\node[slack, label={[font=\scriptsize, right=2pt]{$0$}}] (n0) at (0,2.4) {};
\node[bus, label={[font=\scriptsize, left=2pt]{$i$}}] (ni) at (0,1.1) {};
\node[bus, label={[font=\scriptsize, left=2pt]{$j$}}] (nj) at (0,-0.3) {};
\node[bus, label={[font=\scriptsize, left=2pt]{$k$}}] (nk) at (0,-1.6) {};

\draw[feeder] (grid.south) -- (n0) -- (ni) -- (nj) -- (nk);
\draw[feeder, densely dotted] (nk) -- (0,-2.5);
\node[font=\tiny, gray] at (0.45,-2.35) {\ldots};

\node[callout] at (0.85,1.75) {$P_{0i,t},Q_{0i,t}$};
\node[callout, rotate=90, anchor=center] at (-0.55,0.4) {%
  $v_{i,t}\!\to\!v_{j,t}$\\[-1pt]
  {\tiny LinDistFlow}};

\node[derbox, fill=gray!8] (load) at (2.6,1.1) {Load\\$p^{\mathrm{load}}_{i,t}$};
\draw[feeder, ->] (ni) -- (load.west);

\node[derbox, fill=yellow!15, minimum width=1.55cm] (netder) at (2.9,-0.3) {%
  Net DER\\$p^{\mathrm{DER}}_{j,t}$};
\draw[feeder] (nj) -- (netder.west);

\node[callout, draw=blue!55!black, densely dashed] (doe) at (4.85,0.75) {%
  DOE\\[-1pt]
  $d_{j,t}^{\mathrm{tech}},\, d_{j,t}^{\mathrm{fair}}$};
\draw[thin, densely dashed, blue!55!black] (doe.south west) -- (netder.north east);

\node[derbox, fill=orange!12] (bess) at (5.05,-0.3) {%
  BESS\\$p^{\mathrm{ch}}_{j,t},\, p^{\mathrm{dis}}_{j,t}$};
\draw[feeder, <->] (netder.east) -- (bess.west);

\node[derbox, fill=cyan!10] (wind) at (2.0,-2.0) {Wind\\$p^{\mathrm{W}}_{j,t}$};
\node[derbox, fill=green!10] (pv) at (3.8,-2.0) {PV\\$p^{\mathrm{PV}}_{j,t}$};

\draw[feeder, <-] (netder.south) -- ++(0,-0.35) -| (wind.north);
\draw[feeder, <-] (netder.south) -- ++(0,-0.35) -| (pv.north);

\node[callout] at (2.9,-1.35) {$c^{\mathrm{curt}} P_{j,t}^{\mathrm{curt}}$};

\node[anchor=north west, font=\scriptsize, align=left, draw=black!40,
  rounded corners=2pt, inner sep=4pt, fill=white]
  at (2.7,4.0) {
  \textbf{Variable locations}\\[1pt]
  {\color{red!65!black}$\bullet$} import at feeder head\\
  {\color{blue!55!black}$\bullet$} DOE on net export\\
  {\color{green!50!black}$\bullet$} curtailment at PV/wind\\
  {\color{orange!70!black}$\bullet$} charge/discharge at BESS\\
  {\color{black}$\bullet$} $P,Q,v$ on feeder\\
  {\color{black}$\bullet$} load uncontrolled
};

\end{tikzpicture}

%% file: references.bib
@article{lankeshwara2025,
	author = {Lankeshwara, G. and Sharma, R. and Yan, R. and Tushar, W. and Saha, T. K.},
	journal = {Renewable and Sustainable Energy Reviews},
	title = {{A Review on Network-Aware Control of Distributed Energy Resources under the Dynamic Operating Envelopes Framework}},
	volume = {217},
	pages = {115696},
	year = {2025},
	doi = {10.1016/j.rser.2025.115696}
}

@article{wickramasinghe2025,
	author = {Wickramasinghe, A. and Vilathgamuwa, M. and Nourbakhsh, G. and Corry, P.},
	journal = {Modelling},
	title = {{A Review of Dynamic Operating Envelopes: Computation, Applications and Challenges}},
	volume = {6},
	number = {2},
	pages = {29},
	year = {2025},
	doi = {10.3390/modelling6020029}
}

@article{Chatzivasileiadis_2023_MicroFlex,
title = {Micro-flexibility: {Challenges} for power system modeling and control},
author = {Spyros Chatzivasileiadis and Petros Aristidou and Ioannis Dassios and Tomislav Dragicevic and Daniel Gebbran and Federico Milano and Claudia Rahmann and Deepak Ramasubramanian},
journal = {Electric Power Systems Research},
volume = {216},
pages = {109002},
year = {2023},
issn = {0378-7796},
}

@misc{REN21_2025,
author = {{REN 21}},
title = {{Renewables 2024 Global Status Report (GSR)}},
year = {2025}
}

@ARTICLE{Johanna_Mathieu_2025_IEEE_TEM__DR_DER,
  author={Mathieu, Johanna L. and Verbič, Gregor and Morstyn, Thomas and Almassalkhi, Mads R. and Baker, Kyri and Braslavsky, Julio and Bruninx, Kenneth and Dvorkin, Yury and Ledva, Gregory S. and Mahdavi, Nariman and Pandžić, Hrvoje and Parisio, Alessandra and Perić, Vedran S.},
  journal={IEEE Transactions on Energy Markets, Policy and Regulation}, 
  title={A New Definition and Research Agenda for Demand Response in the Distributed Energy Resource Era}, 
  year={2025},
  volume={3},
  number={3},
  pages={324-339},
}

@article{baran1989,
	author = {Baran, M. E. and Wu, F. F.},
	journal = {IEEE Transactions on Power Delivery},
	title = {{Optimal Capacitor Placement on Radial Distribution Systems}},
	volume = {4},
	number = {1},
	pages = {725--734},
	year = {1989},
	doi = {10.1109/61.19265}
}

@article{farivar2013,
	author = {Farivar, M. and Low, S. H.},
	journal = {IEEE Transactions on Power Systems},
	title = {{Branch Flow Model: Relaxations and Convexification---Part I}},
	volume = {28},
	number = {3},
	pages = {2554--2564},
	year = {2013},
	doi = {10.1109/TPWRS.2013.2255317}
}

@article{boyd2011,
	author = {Boyd, S. and Parikh, N. and Chu, E. and Peleato, B. and Eckstein, J.},
	journal = {Foundations and Trends in Machine Learning},
	title = {{Distributed Optimization and Statistical Learning via the Alternating Direction Method of Multipliers}},
	volume = {3},
	number = {1},
	pages = {1--122},
	year = {2011},
	doi = {10.1561/2200000016}
}

@article{peng2018,
	author = {Peng, Q. and Low, S. H.},
	journal = {IEEE Transactions on Smart Grid},
	title = {{Distributed Optimal Power Flow Algorithm for Radial Networks, {I}: Balanced Single Phase Case}},
	volume = {9},
	number = {1},
	pages = {111--121},
	year = {2018},
	doi = {10.1109/TSG.2016.2546305}
}

@article{petrou2021,
	author = {Petrou, K. and Procopiou, A. T. and Ochoa, L. F. and Gutierrez-Lagos, L. and Liu, M. Z. and Langstaff, T. and Theunissen, J.},
	journal = {IEEE Transactions on Smart Grid},
	title = {{Ensuring Distribution Network Integrity Using Dynamic Operating Limits for Prosumers}},
	volume = {12},
	number = {5},
	pages = {3877--3888},
	year = {2021},
	doi = {10.1109/TSG.2021.3081371}
}

@misc{jain1984,
	author = {Jain, R. and Chiu, D. and Hawe, W.},
	title = {{A Quantitative Measure of Fairness and Discrimination for Resource Allocation in Shared Computer Systems}},
	howpublished = {DEC Research Report TR-301},
	month = {Sep.},
	year = {1984},
	note = {Also available as arXiv:cs/9809099}
}

@article{thurner2018,
	author = {Thurner, L. and Scheidler, A. and Sch{\"a}fer, F. and Menke, J.-H. and Dollichon, J. and Meier, F. and Meinecke, S. and Braun, M.},
	journal = {IEEE Transactions on Power Systems},
	title = {{pandapower---An Open-Source {Python} Tool for Convenient Modeling, Analysis, and Optimization of Electric Power Systems}},
	volume = {33},
	number = {6},
	pages = {6510--6521},
	year = {2018},
	doi = {10.1109/TPWRS.2018.2829021}
}

@article{yi2022,
	author = {Yi, Y. and Verbi{\v{c}}, G.},
	journal = {Electric Power Systems Research},
	title = {{Fair Operating Envelopes under Uncertainty Using Chance Constrained Optimal Power Flow}},
	volume = {213},
	pages = {108465},
	year = {2022},
	doi = {10.1016/j.epsr.2022.108465}
}

@article{gebbran2021,
	author = {Gebbran, D. and Mhanna, S. and Ma, Y. and Chapman, A. C. and Verbi{\v{c}}, G.},
	journal = {Applied Energy},
	title = {{Fair Coordination of Distributed Energy Resources with Volt-Var Control and {PV} Curtailment}},
	volume = {286},
	pages = {116546},
	year = {2021},
	doi = {10.1016/j.apenergy.2021.116546}
}

@INPROCEEDINGS{gupta2024,
	author = {Gupta, R. K. and Buason, P. and Molzahn, D. K.},
	booktitle = {Proc. 2024 IEEE Texas Power and Energy Conference (TPEC)},
	title = {{Fairness-Aware Photovoltaic Generation Limits for Voltage Regulation in Power Distribution Networks Using Conservative Linear Approximations}},
	year = {2024},
	pages = {1--6},
	doi = {10.1109/TPEC60005.2024.10472277}
}

@article{pinto2020,
	author = {Pinto, R. and Marcelino, C. G. and Heleno, M. and Mancarella, P.},
	journal = {Electric Power Systems Research},
	title = {{Distributed Multi-Period Three-Phase Optimal Power Flow Using Temporal Neighbors}},
	volume = {182},
	pages = {106228},
	year = {2020},
	doi = {10.1016/j.epsr.2020.106228}
}

@article{molzahn2017,
	author = {Molzahn, D. K. and D{\"o}rfler, F. and Sandberg, H. and Low, S. H. and Chakrabarti, S. and Baldick, R. and Lavaei, J.},
	journal = {IEEE Transactions on Smart Grid},
	title = {{A Survey of Distributed Optimization and Control Algorithms for Electric Power Systems}},
	volume = {8},
	number = {6},
	pages = {2941--2962},
	year = {2017},
	doi = {10.1109/TSG.2017.2720471}
}

@INPROCEEDINGS{bernstein2017,
	author = {Bernstein, A. and Dall'Anese, E.},
	booktitle = {Proc. 2017 IEEE PES Innovative Smart Grid Technologies Conference Europe (ISGT-Europe)},
	title = {{Linear Power-Flow Models in Multiphase Distribution Networks}},
	year = {2017},
	pages = {1--6},
	doi = {10.1109/ISGTEurope.2017.8260205}
}

@INPROCEEDINGS{li2011,
	author = {Li, N. and Chen, L. and Low, S. H.},
	booktitle = {Proc. IEEE Power and Energy Society General Meeting},
	title = {{Optimal Demand Response Based on Utility Maximization in Power Networks}},
	year = {2011},
	pages = {1--8},
	doi = {10.1109/PES.2011.6039082}
}

@article{zamzam2018,
	author = {Zamzam, A. S. and Sidiropoulos, N. D. and Dall'Anese, E.},
	journal = {IEEE Transactions on Smart Grid},
	title = {{Beyond Relaxation and Newton--Raphson: Solving {AC} {OPF} for Multi-Phase Systems with Renewables}},
	volume = {9},
	number = {5},
	pages = {3966--3975},
	year = {2018},
	doi = {10.1109/TSG.2016.2645220}
}

@article{liu2022oe,
	author = {Liu, M. Z. and Ochoa, L. F. and Wong, P. K. C. and Theunissen, J.},
	journal = {IEEE Transactions on Smart Grid},
	title = {{Using {OPF}-Based Operating Envelopes to Facilitate Residential {DER} Services}},
	volume = {13},
	number = {6},
	pages = {4494--4504},
	year = {2022},
	doi = {10.1109/TSG.2022.3188927}
}

@INPROCEEDINGS{petrou2020fair,
	author = {Petrou, K. and Liu, M. Z. and Procopiou, A. T. and Ochoa, L. F. and Theunissen, J. and Harding, J.},
	booktitle = {Proc. 2020 IEEE PES Innovative Smart Grid Technologies Europe (ISGT-Europe)},
	title = {{Operating Envelopes for Prosumers in {LV} Networks: A Weighted Proportional Fairness Approach}},
	year = {2020},
	pages = {579--583},
	doi = {10.1109/ISGT-Europe47291.2020.9248975}
}

@article{alam2023,
	author = {Alam, M. R. and Nguyen, P. T. H. and Naranpanawe, L. and Saha, T. K. and Lankeshwara, G.},
	journal = {IEEE Transactions on Sustainable Energy},
	title = {{Allocation of Dynamic Operating Envelopes in Distribution Networks: Technical and Equitable Perspectives}},
	volume = {15},
	number = {1},
	pages = {173--186},
	year = {2024},
	doi = {10.1109/TSTE.2023.3275082}
}

@article{gebbran2023,
	author = {Gebbran, D. and Mhanna, S. and Chapman, A. C. and Verbi{\v{c}}, G.},
	journal = {IEEE Transactions on Smart Grid},
	title = {{Multiperiod {DER} Coordination Using {ADMM}-Based Three-Block Distributed {AC} Optimal Power Flow Considering Inverter Volt-Var Control}},
	volume = {14},
	number = {4},
	pages = {2874--2889},
	year = {2023},
	doi = {10.1109/TSG.2022.3227635}
}

@INPROCEEDINGS{Gebbran_SGES,
author={Gebbran, Daniel and Mhanna, Sleiman and Chapman, Archie C. and Hardjawana, Wibowo and Vucetic, Branka and Verbi\v{c}, Gregor},
booktitle={Proceedings of the 2020 International Conference on Smart Grids and Energy Systems (SGES 2020)},
title={Practical Considerations of {DER} Coordination with Distributed Optimal Power Flow},
year={2020},
pages={1-6},
month={Nov.}}

@article{Lubin2023,
	author = {Lubin, Miles and Dowson, Oscar and Dias Garcia, Joaquim and Huchette, Joey and Legat, Beno{\^i}t and Vielma, Juan Pablo},
	title = {{JuMP} 1.0: Recent improvements to a modeling language for mathematical optimization},
	journal = {Mathematical Programming Computation},
	volume = {15},
	pages = {581--589},
	year = {2023},
	doi = {10.1007/s12532-023-00239-3}
}

@article{Huangfu2018,
	author = {Huangfu, Q. and Hall, J. A. J.},
	title = {Parallelizing the dual revised simplex method},
	journal = {Mathematical Programming Computation},
	volume = {10},
	number = {1},
	pages = {119--142},
	year = {2018},
	doi = {10.1007/s12532-017-0130-5}
}
